\documentclass[12pt]{article}
\begin{document}
\vbadness = 100000
\hbadness = 10000
\newcommand{\grad}{\mbox{\boldmath$\nabla$}}
\newcommand{\bdiv}{\mbox{\boldmath$\nabla\cdot$}}
\newcommand{\curl}{\mbox{\boldmath$\nabla\times$}}
\newcommand{\bcdot}{\mbox{\boldmath$\cdot$}}
\newcommand{\btimes}{\mbox{\boldmath$\times$}}
\newcommand{\btau}{\mbox{\boldmath$\tau$}}
\newcommand{\btheta}{\mbox{\boldmath$\theta$}}
\newcommand{\bphi}{\mbox{\boldmath$\phi$}}
\newcommand{\bmu}{\mbox{\boldmath$\mu$}}
\newcommand{\bepsilon}{\mbox{\boldmath$\epsilon$}}
\newcommand{\bcj}{\mbox{\boldmath$\cal J$}}
\newcommand{\bcf}{\mbox{\boldmath$\cal F$}}
\newcommand{\bbeta}{\mbox{\boldmath$\beta$}}
\newcommand{\lbar}{\lambda\hspace{-.09in}^-}
\newcommand{\bcp}{\mbox{\boldmath$\cal P$}}
\newcommand{\bco}{\mbox{\boldmath$\omega$}}
\title{The electromagnetic momentum of static charge-current distributions}
\author{Jerrold Franklin\footnote{Internet address:
Jerry.F@TEMPLE.EDU}\\
Department of Physics\\
Temple University, Philadelphia, PA 19122-6082
\date{\today}}
\maketitle
\begin{abstract}
The origin of electromagnetic momentum for general static charge-current distributions is examined.  
The electromagnetic momentum for static electromagnetic fields is derived by implementing conservation of momentum for the sum of mechanical momentum and electromagnetic momentum. The external force required to keep matter at rest during the production of the 
final static configuration produces the electromagnetic momentum.
Examples of the electromagnetic momentum in static electric and magnetic fields are given.  
The `center of energy' theorem is shown to be violated by electromagnetic momentum.
`Hidden momentum' is shown to be generally absent, and not to cancel electromagnetic momentum.
\end{abstract}

\section{Electromagnetic momentum}
The momentum of static electromagnetic  fields has been discussed in many papers with inconsistent results.
A recent review article\cite{dgl} lists 143 references for electromagnetic momentum. Thus many of the equations in this paper will have appeared before, but we will often have somewhat different derivations and new observations.
 
Electromagnetic (EM) fields exert forces on matter via the Lorentz force (in Gaussian units)
\begin{equation}
{\bf F}_{\rm Lorentz}=\int\left(\rho{\bf E}+\frac{1}{c}{\bf j\times B}\right)d^3r.
\label{eq:fmat}
\end{equation}
In the absence of any other force on the matter,
\begin{equation}
{\bf F}_{\rm Lorentz}=\frac{\bf dP_{\rm Matter}}{dt}.
\label{eq:fm}
\end{equation}
With no external force, the introduction of EM momentum is needed to preserve conservation of momentum.  That is,
\begin{equation}
\frac{\bf dP_{\rm Matter}}{dt}+\frac{\bf dP_{\rm EM}}{dt}=0
\label{eq:n3}
\end{equation}
expresses the law of conservation of momentum.
Then, the rate of change of EM momentum is given by
\begin{equation}
\frac{\bf dP_{\rm EM}}{dt}=-\int\left(\rho{\bf E}+\frac{1}{c}{\bf j\times B}\right)d^3r.
\label{eq:dpdtem}
\end{equation}

Equations (\ref{eq:fmat}) and (\ref{eq:fm}) describe the motion of matter that is caused by the EM fields with no other forces acting.  We are interested in the case where an external force acts directly on the matter to hold it in place.  Then,
\begin{equation}
{\bf F}_{\rm External}=-{\bf F}_{\rm Lorentz}=\frac{\bf dP_{\rm EM}}{dt}.
\label{eq:n12}
\end{equation}
The external force keeps the matter in place while increasing the EM momentum.
Previous discussions of the production of EM momentum have generally left out the important role of the external force in keeping the matter at rest\cite{b1}.

The preceding discussion has made use of all three of Newton's basic laws of motion, suitably framed in terms of momentum:
\begin{itemize}
\item Equation (\ref{eq:n3}) expresses Newton's third law (in terms of momentum), with the rate of change of EM momentum being equal and opposite to the rate of change of the momentum of the matter on which the EM fields act.  With the inclusion of EM momentum, Newton's third law still holds but must be formulated in terms of conservation of overall momentum. 
\item The first equality of Eq.~(\ref{eq:n12}) expresses Newton's first law, that matter will remain at rest if acted on by equal and opposite forces.
\item The second equality of Eq.~(\ref{eq:n12}) expresses Newton's second law that the external force equals the rate of change of overall momentum, 
which in this case is the EM momentum. 
\end{itemize}

Equation (\ref{eq:n12}) shows that an external force will increase the EM momentum while keeping matter at rest.  This is a counter-example to a proposed `center of energy' theorem that  states\cite{dgl} ``If the center of energy of a closed system is at rest, then its total momentum is zero." Attempts to reconcile the center of energy theorem with the presence of EM momentum have led to the introduction of  `hidden momentum' to cancel the EM momentum, with the claim that a body can have this hidden momentum without moving.  We show  in Section 6 of this paper
how hidden momentum is absent, and the center of energy theorem is violated, in the generation of EM momentum.  
Simply put, the external force has produced momentum that has gone into the EM field.  If the total momentum (EM momentum plus hidden momentum) were now zero, then the momentum  produced by the external force would have been lost.

\section{EM momentum of static charge-current distributions}

Here we derive the EM momentum of two distinct distributions, a charge density $\rho_1({\bf r})$ and a current density ${\bf j}_2{\bf(r)}$.
We use two distinct distributions so as to avoid any self forces that could confuse the situation.  
We start with a situation where only the charge distribution $\rho_1$ is present, with no initial EM momentum.
Then we introduce a time dependent current distribution ${\bf\overline j}_2({\bf r},t)$ that increases from zero to its final value ${\bf j}_2({\bf r})$,
after which it remains constant in time\cite{tdj}.
 
We integrate Eq.~(\ref{eq:dpdtem}) over time to get the final static EM momentum
\begin{equation}
{\bf P}_{\rm EM}=\int \frac{\bf dP_{\rm EM}}{dt}dt=-\int dt\int\rho_1({\bf r){\overline E}_2( r},t) d^3r.
\label{eq:pem1}
\end{equation}
There is no force on the current distribution because the static charge distribution $\rho_1$ produces no magnetic field,
and its static electric field exerts no force on the current distribution.

The time dependent current distribution will produce a time dependent vector potential, and an electric field given by
\begin{equation}
{\bf{\overline E}}_2{\bf(r},t)=-\frac{1}{c}\partial_t{\bf{\overline A}}_2({\bf r},t).
\label{ead}
\end{equation}
Then
\begin{equation}
{\bf P}_{\rho{\bf A}}=\frac{1}{c}\int dt\int\rho_1({\bf r})\partial_t{\bf{\overline A}}_2({\bf  r},t) d^3r=\frac{1}{c}\int\rho_1({\bf r}){\bf A}_2({\bf  r}) d^3r.
\label{eq:pemra}
\end{equation}
We have labeled this form of the EM momentum as ${\bf P}_{\rho{\bf A}}$ to distinguish it from other forms we will be deriving. 

It is important to note that Eq.~(\ref{eq:pemra}) will hold only if an external force has acted while the current was increasing to its final value.  Without this external force, the charge distribution $\rho_1$ would not have remained at rest.  This means that the final state of static charge-current distributions and EM fields has the momentum produced by the external force.    

We can use the law of Biot-Savart to
write ${\bf P}_{\rm EM}$ in terms of $\rho_1$ and ${\bf j}_2$ as
\begin{equation}
{\bf P}_{\rho{\bf j}}=\frac{1}{c^2}\int d^3r\int  d^3r'\frac{\rho_1({\bf r}){\bf j}_2({\bf  r'})}{|{\bf r-r'}|}.
\label{eq:pembs}
\end{equation}
Now we can apply Coulomb's law to Eq.~(\ref{eq:pembs})
to get another form of ${\bf  P}_{\rm EM}$ as
\begin{equation}
{\bf P}_{\phi{\bf j}}=\frac{1}{c^2}\int\phi_1({\bf r}){\bf j}_2({\bf  r}) d^3r.
\label{eq:pempj}
\end{equation}
It is of interest to note that that Eq.~(\ref{eq:pempj}) seems to suggest that the EM momentum resides in the current distribution, while Eq.~(\ref{eq:pemra}) seems to suggest that the EM momentum resides in the charge distribution.  The resolution is that Eq.~(\ref{eq:pembs}) shows that ${\bf P}_{\rm EM}$ is due to the cooperative interaction of both distributions, with neither being dominant.

We can apply Maxwell's equations to Eq.~(\ref{eq:pempj}) to get a form of ${\bf P}_{\rm EM}$ that depends only on the {\bf E} and {\bf B} fields, 
\begin{eqnarray}
{\bf P}_{\bf EBj}&=&\frac{1}{c^2}\int\phi_1{\bf j}_2 d^3r\nonumber\\
&=&\frac{1}{4\pi c}\int\phi_1({\bf\curl B}_2)d^3r\nonumber\\
&=&\frac{1}{4\pi c}\int[{\bf\curl( B}_2\phi_1)+{\bf B}_2\times\grad\phi_1]d^3r\nonumber\\
&=&\frac{1}{4\pi c}\int_V{\bf( E}_1\times{\bf B}_2)d^3r
+\frac{1}{4\pi c}\oint_S{\bf dS\times B}_2\phi_1.
\label{eq:pemebj}
\end{eqnarray}
This form of ${\bf P}_{\rm EM}$, without the surface integral, is the form that is usually given in EM texts.
However, if the surface integral in Eq.~(\ref{eq:pemebj}) does not vanish, it must be included to give the total EM momentum.

Using Eq.~(\ref{eq:pemebj}), an EM momentum density can be defined by
\begin{equation}
{\bf g}_{\rm EM}=\frac{1}{4\pi c}({\bf E\times B)}.
\end{equation}
Then the integral of ${\bf g}_{\rm EM}$ over any volume, without adding the surface integral, gives the EM momentum within that volume.
The location of the EM momentum is thus given by  ${\bf g}_{\rm EM}$.  It does not reside solely in either the charge or the current distribution. 
If the surface integral in Eq.~(\ref{eq:pemebj}) is included, the result is the total EM momentum of all charge distributions within the volume.
Including the surface integral accounts for EM momentum that left the volume during the production of the EM momentum.

An alternate derivation of the ($\bf E\times B$) form starts from Eq.~(\ref{eq:pemra}),
\begin{eqnarray}
{\bf P}_{\bf EB\rho}&=&\frac{1}{c}\int\rho_1({\bf r}){\bf A}_2({\bf  r}) d^3r\nonumber\\
&=&\frac{1}{4\pi c}\int{\bf A}_2(\grad\cdot{\bf E}_1) d^3r\nonumber\\
&=&\frac{1}{4\pi c}\int[\grad\cdot({\bf E}_1{\bf A}_2)-({\bf E}_1\cdot\grad){\bf A}_2]d^3r\nonumber\\
&=&\frac{1}{4\pi c}\int[\grad\cdot({\bf E}_1{\bf A}_2)
+{\bf E}_1\times(\curl{\bf A}_2)-\grad({\bf E}_1\cdot{\bf A}_2)\nonumber\\
&&\hspace{.5in}+({\bf A}_2\cdot\grad){\bf E}_1+{\bf A}_2\times(\curl{\bf E}_1)]d^3r\nonumber\\
&=&\frac{1}{4\pi c}\int[\grad\cdot({\bf E}_1{\bf A}_2)+ ({\bf E}_1\times{\bf B}_2)-\grad({\bf E}_1\cdot{\bf A}_2)+\grad\cdot({\bf A}_2{\bf E}_1)]d^3r\nonumber\\
\hspace{.5in}
&=&\frac{1}{4\pi c}\int_V{\bf( E}_1\times{\bf B}_2)d^3r
\nonumber\\&&
+\frac{1}{4\pi c}\oint_S{\bf dS\cdot[(E}_1{\bf A}_2+{\bf A}_2{\bf E}_1
-{\hat{\hat{\bf n}}}({\bf E}_1\cdot{\bf A}_2)],
\label{eq:pemebr}
\end{eqnarray}
where ${\hat{\hat{\bf n}}}$ is the unit dyadic.
In this derivation we used the facts that the  curl of $\bf E$ and the divergence of $\bf A$ each vanish for static distributions.

If the volume integral includes only the charge distribution $\rho_1$, Eq.~(\ref{eq:pemebr})  must be used with its surface integral to get the total
EM momentum.
If the volume integral includes only the current distribution ${\bf j}_2$, Eq.~(\ref{eq:pemebj})  must be used with its surface integral.
We will give examples of each case in the next section.
If the volume integral includes both the charge and the current distributions, either equation may be used.  In that case, the two surface integrals turn out to be equal,
and will usually go to zero for an infinite volume.

One more form for the EM momentum can be derived, starting from Eq.~(\ref{eq:pempj})
\begin{eqnarray}
{\bf P}_{\bf rjE}&=&\frac{1}{c^2}\int\phi_1{\bf j}_2 d^3r\nonumber\\
&=&\frac{1}{c^2}\int\phi_1({\bf j}_2\cdot\grad){\bf r} d^3r\nonumber\\
&=&\frac{1}{c^2}\int[\grad\cdot({\bf j}_2{\bf r}\phi_1) 
-{\bf rj}_2\cdot(\grad\phi_1)-\phi_1{\bf r}(\grad\cdot{\bf j}_2)]d^3r\nonumber\\
&=&\frac{1}{c^2}\int{\bf r(j}_2\cdot{\bf E}_1)d^3r
+\frac{1}{c^2}\oint{\bf dS\cdot j}_2{\bf r}\phi_1\nonumber\\
&=&\frac{1}{c^2}\int{\bf rj}_2\cdot{\bf E}_1d^3r.
\label{eq:pemrje}
\end{eqnarray}
In this derivation we used the fact that $\grad\cdot{\bf j}=0$, and that the surface integral vanishes if no current passes through the surface.

\section{EM momentum of an magnetic dipole in  a static electric field}

In the previous section, we derived six equivalent expressions for the EM momentum of static charge-current distributions.
In this section, we will apply those results to find the EM momentum of a magnetic dipole in a static electric field.
We will first use the simplest of the forms, but (with a bit more complication) show that the other forms give the same result.
For the two $\bf E\times B$ forms, the appropriate surface integral has to be included if it does not vanish.

For a magnetic dipole in an electric field, the simplest form to use is that of Eq.~(\ref{eq:pemra}), with 
\begin{equation}
{\bf A}=\frac{\bmu\times{\bf r}}{r^3}
\label{eq:a}
\end{equation}
for the vector potential of a magnetic dipole.
This gives
\begin{eqnarray}
{\bf P}_{\rm EM}&=&\frac{1}{c}\int\rho{\bf A} d^3r\nonumber\\
&=&\frac{1}{c}\int\frac{\rho\bmu\times{\bf r}}{r^3}d^3r\nonumber\\
&=&\frac{1}{c}{\bf E\times\bmu},
\label{eq:edip}
\end{eqnarray}
where $\bf E$ is the electric field at the position of the dipole.
We have reversed the sign on $\bf E$ in the last step above, because the $\bf r$ in the previous line points from the dipole toward the charge density.

Interestingly, this result for the EM momentum of a magnetic dipole in a static electric field
 does not depend on the form of the magnetic dipole, but only on the fact that the vector potential is 
given by Eq.~(\ref{eq:a}).  Neither does it depend on the nature of the source of the static electric field.
Thus, the result is the same for the electric field from a point charge, for which 
\begin{equation}
{\bf P}_{\rm EM}=\frac{q\bf{\hat r}\times\bmu}{cr^2},
\label{eq:qmu}
\end{equation}
an electric dipole, for which
\begin{equation}
{\bf P}_{\rm EM}=\frac{[3(\bf p\cdot{\hat r})({\hat r}\times\bmu)-p\times\bmu]}{cr^3},
\label{eq:pmu}
\end{equation}
a uniform $\bf E$ field, or any other kind of static electric field.
Furry\cite{wf} found the result of Eq.~(\ref{eq:edip}) for the special case of a point charge and magnetic moment, with a more complicated derivation using the 
($\bf E\times B$) form.  

As shown by our derivation in section 2, the other forms for ${\bf P}_{\rm EM}$ should all give the same result as Eq.~(\ref{eq:edip}),
however a straightforward derivation using those forms is only possible for a  uniform electric field.
In Eq.~(\ref{eq:pempj}),
\begin{equation}
{\bf P}_{\phi{\bf j}}=\frac{1}{c^2}\int\phi({\bf r}){\bf j}({\bf  r}) d^3r,
\label{eq:pempj2}
\end{equation}
we can take 
\begin{equation}
\phi({\bf r)=-r\cdot E},
\label{eq:pre}
\end{equation}
which holds for a uniform electric field.
Using Eq.~(\ref{eq:pre}) in Eq.~(\ref{eq:pempj2}) gives
\begin{equation}
{\bf P}_{\rm EM}=-\frac{1}{c^2}\int d^3r{\bf j}{\bf  r} \cdot{\bf E}.
\label{eq:pempj3}
\end{equation}
This form  is similar to Eq.~(\ref{eq:pemrje}), and either
integral can be related to the magnetic moment
following the usual textbook\cite{jf,dj} derivation. 
We write the integral in Eq.~(\ref{eq:pempj3}) as
\begin{eqnarray}
{\bf P}_{\rm EM}&=&-\frac{1}{c^2}\int{\bf jr\cdot E}d^3r\nonumber\\
&=&-\frac{1}{2c^2}\int[{\bf(jr+rj)+(jr-rj)]\cdot E}d^3r.
\label{eq:calk2}
\end{eqnarray}
The first, symmetric combination can be converted into a divergence (using $\grad\cdot{\bf j}=0$), and then into a surface integral that vanishes if taken outside the current distribution.  That leaves
\begin{eqnarray}\
{\bf P}_{\rm EM}&=&\frac{1}{2}\int{\bf[r(j\cdot E)-j(r\cdot E)]}d^3r\nonumber\\
&=&\frac{1}{2c^2}\int[{\bf E\times(r\times j)}]d^3r\nonumber\\
&=&\frac{1}{c}{\bf E\times\bmu},
\label{eq:calkp}
\end{eqnarray}
where we have recognized
\begin{equation}
\bmu=\frac{1}{2c}\int({\bf r\times j})d^3r
\label{eq:mrj}
\end{equation}
as a representation of the magnetic moment.

To use the $({\bf E\times B})$ form for ${\bf P}_{\rm EM}$ for a magnetic dipole in a uniform electric field, we use Eq.~({\ref{eq:pemebj}) and integrate over a sphere of radius $R$  about the magnetic dipole.
The dipole's magnetic field is
\begin{equation}
{\bf B(r)}= \frac{3{\bf(\bmu\bcdot\hat{r})\hat{r}-\bmu}}{r^3}
+4\pi{\bf[\bmu-\hat{r}(\hat{r}\bcdot \bmu)}]\delta({\bf r}).
\label{eq:sing2}
\end{equation}
With this form for $\bf B(r)$, the volume integral in Eq.~({\ref{eq:pemebj}) gives
\begin{eqnarray}
I_V&=&\frac{1}{4\pi c}\int_V{\bf( E\times B)}d^3r\nonumber\\
&=&\frac{1}{4\pi c}{\bf  E}\times\int_V {\bf B}d^3r\nonumber\\
&=&-\frac{1}{4\pi c}{\bf  E}\times\left\{\int_V \frac{[3{\bf(\bmu\bcdot\hat{r})\hat{r}-\bmu}]}{r^3}d^3r
-4\pi\int_V[\bmu-{\bf\hat{r}(\hat{r}\bcdot\bmu)}]\delta({\bf r})d^3r\right\}\nonumber\\
&=&\frac{2}{3c}{\bf(E\times\bmu)},
\end{eqnarray}
where the result comes from the angular integration of the singular part of the magnetic field
since the angular integration of the first integral vanishes.
The surface integral gives
\begin{eqnarray}
I_S&=&\frac{1}{4\pi c}\oint_S{\bf dS\times B\phi}\nonumber\\
&=&-\frac{1}{4\pi c}\oint d\Omega{\bf{\hat r}}
\times  [3{\bf(\bmu\bcdot\hat{r})\hat{r}-\bmu}]({\bf{\hat r}\cdot E})\nonumber\\
&=&\frac{1}{3c}{\bf(E\times\bmu)}.
\end{eqnarray}

We see that the sum of the volume and surface integrals gives the correct answer, ${\bf E\times\bmu}/c$,
but leaving out the surface integral would give a misleading, incorrect result.
The surface integral is independent of the radius of the sphere, so even integrating over `all space' requires the surface integral contribution.
This is true as long as the integration volume, no matter how large, does not include the much larger parallel plates producing the uniform electric field.

\section{EM momentum of an electric dipole in  a uniform magnetic field}

We will see in each derivation below 
that we can only get a simple form for
the EM momentum of an electric dipole moment $\bf p$ in a magnetic field $\bf B$ if the field is uniform.  
The simplest derivation uses Eq.~(\ref{eq:pemra}) with ${\bf A=(B\times r)}/2$ for a uniform $\bf B$ field.
Then we have
\begin{eqnarray}
{\bf P}_{\rm EM}&=&\frac{1}{c}\int\rho{\bf A} d^3r\nonumber\\
&=&\frac{1}{2c}\int\rho({\bf B\times r})d^3r\nonumber\\
&=&\frac{1}{2c}{\bf B\times p}.
\label{eq:pedB}
\end{eqnarray}
We note again that this result is only valid for a uniform $\bf B$ field.  For instance, if we put the magnetic field of a magnetic dipole into Eq.~(\ref{eq:pedB}), it would give the wrong result for ${\bf P}_{\rm EM}$ of an electric and magnetic dipole, which is given correctly by Eq.~(\ref{eq:pmu}) above.

To use the $({\bf E\times B})$ form for ${\bf P}_{\rm EM}$ of an electric dipole in a uniform magnetic field,
we use Eq.~({\ref{eq:pemebr}) and integrate over a sphere of radius $R$  about the electric dipole, 
whose electric field is
\begin{equation}
{\bf E(r)}= \frac{3{\bf(p\bcdot\hat{r})\hat{r}-p}}{r^3}
-4\pi{\bf\hat{r}(\hat{r}\bcdot p)}\delta({\bf r}).
\label{eq:singe}
\end{equation}
With this form of $\bf E(r)$, the volume integral in Eq.~({\ref{eq:pemebr}) gives
\begin{eqnarray}
I_V&=&\frac{1}{4\pi c}\int_V{\bf( E\times B)}d^3r\nonumber\\
&=&-\frac{1}{4\pi c}{\bf  B}\times\int_V {\bf E}d^3r\nonumber\\
&=&-\frac{1}{4\pi c}{\bf  B}\times\left\{\int_V \frac{[3{\bf(p\bcdot\hat{r})\hat{r}-p}]}{r^3}d^3r
-4\pi\int_V{\bf\hat{r}(\hat{r}\bcdot p)}\delta({\bf r})d^3r\right\}\nonumber\\
&=&\frac{1}{3c}{\bf(B\times p)},
\end{eqnarray}
where the result comes from the angular integration of the singular part of the electric field since the angular integration of the first integral vanishes.

The surface integral in Eq.~({\ref{eq:pemebr}) gives
\begin{eqnarray}
I_S&=&\frac{1}{4\pi c}\oint_S{\bf dS\cdot[(E}{\bf A}+{\bf A}{\bf E})+
{\hat{\hat{\bf n}}}({\bf E}\cdot{\bf A})]\nonumber\\
&=&-\frac{1}{8\pi c}\oint d\Omega{\bf{\hat r}}\cdot\{ [3{\bf(p\cdot{\hat r}){\hat r}-p}]
{\bf(B\times{\hat r})}+{\bf(B\times{\hat r})}[3{\bf(p\cdot{\hat r}){\hat r}-p}]\nonumber\\
&&+{\hat{\hat{\bf n}}}[3{\bf(p\cdot{\hat r}){\hat r}-p}]\cdot{\bf(B\times{\hat r}})\}\nonumber\\
&=&\frac{1}{6c}{\bf(B\times\ p)}.
\end{eqnarray}
We see that the sum of the volume and surface integrals gives the correct answer, ${\bf B\times p}/2c$,
but leaving out the surface integral would give a misleading, incorrect result.
The surface integral is independent of the radius of the sphere, so even integrating over `all space' requires the surface integral contribution.
This is true as long as the integration volume, no matter how large, does not include the much larger solenoid producing the uniform magnetic field at its center.

So far, we have produced the final static charge-current distribution by starting with a fixed charge distribution, and then increasing the current distribution to its final value to get  ${\bf P}_{\rm EM}$ as we did in Eq.~(\ref{eq:pem1}).
Now we show that the same result for 
${\bf P}_{\rm EM}$ follows if we start with a fixed current distribution, and use a separate current to produce the electric dipole moment.

The final configuration will be an electric dipole
$\bf p$ at the center of a long circular solenoid which produces a constant magnetic field $\bf B$ at the position of the electric dipole.
We can achieve this by starting with a constant current I in the solenoid. We then produce the electric dipole with a current I$'$ in a short length $\bf L$ of wire at the ultimate position of the dipole.  The force on the short current length is given by 
\begin{equation}
{\bf F}_{\bf L}=\frac{II'}{c^2}\int_{-{\bf L}/2}^{+{\bf L}/2}
{\bf dr'}\times\oint_{c}\frac{{\bf[dr}\times{\bf (r'-r)]}}{|{\bf r'-r}|^3}
\label{fbs}
\end{equation}
where the contour c consists of a sum over all the current loops of the solenoid.
We can expand the triple cross product to give
\begin{equation}
{\bf F}_{\bf L}=\frac{II'}{c^2}\left[\oint_{c}
{\bf dr}\int_{-{\bf L}/2}^{+{\bf L}/2}
\frac{{\bf dr'}\cdot{\bf (r'-r)}}{\bf|r'-r|^3}
-\int_{-{\bf L}/2}^{+{\bf L}/2}{\bf dr'}\cdot\oint_{c}\frac{\bf dr(r'-r)}{{\bf|r'-r|}^3}\right].
\label{fonl}
\end{equation}

It is also important to include the force on the solenoid caused by the current in the short length $\bf L$. This is given by
\begin{equation}
{\bf F}_{S}=\frac{II'}{c^2}\left[\int_{-{\bf L}/2}^{+{\bf L}/2}{\bf dr'}\oint_{c}
\frac{{\bf dr}\cdot{\bf (r-r')}}{{\bf|r-r'|}^3}
-\oint_{c}{\bf dr}\cdot\int_{-{\bf L}/2}^{+{\bf L}/2}\frac{\bf dr'(r-r')}{{\bf|r-r'|}^3}\right].
\label{fons}
\end{equation}
The net force on the combined system of the short current length and the solenoid is given by the sum of the two forces. 
In this sum, the second terms in Eqs.~(\ref{fonl}) and (\ref{fons}) cancel, while the first-term in Eq. (\ref{fons}) vanishes because it involves an integral of a gradient over a closed contour.  This leaves
\begin{equation}
{\bf F}_{\bf L}+{\bf F}_S
=\frac{II'}{c^2}\oint_c
{\bf dr}\int_{-{\bf L}/2}^{+{\bf L}/2}
\frac{{\bf dr'}\cdot{\bf (r'-r)}}{{\bf|r'-r|}^3}.
\label{fnet}
\end{equation}

	As in Eq.~(\ref{eq:n12}), this net force must be countered by an external applied force to keep the solenoid and short wire at rest.  This external force produces the EM momentum of the configuration, such that 
\begin{equation}
\frac{\bf dP_{\rm EM}}{dt}=
{\bf F}_{\rm External}=-[{\bf F}_{\bf L}+{\bf F}_{S}]
=-\frac{II'}{c^2}\left[\oint_{c}
{\bf dr}\int_{-{\bf L}/2}^{+{\bf L}/2}\frac{{\bf dr'}\cdot\bf (r-r')}{\bf|r-r'|^3}\right].
\label{eq:n12b}
\end{equation}
For L$<<$R, the radius of the solenoid, we can neglect the $\bf r'$ in the integral over $\bf dr'$. This leaves
\begin{equation}
\frac{\bf dP_{\rm EM}}{dt}=
=-\frac{II'}{c^2}\oint_{c}
{\bf dr}\int_{-{\bf L}/2}^{+{\bf L}/2}\frac{{\bf dr'}\cdot\bf r}{r^3}
=-\frac{II'}{c^2}{\bf L\cdot}\oint_{c}
\frac{{\bf r dr}}{r^3}.
\label{eq:n12c}
\end{equation}

Integrating the current in the short wire over time gives
\begin{equation}
{\bf P}_{\rm EM}=\int\frac{\bf dP_{\rm EM}}{dt}dt=
=-\frac{1}{c^2}q{\bf L}\cdot I\oint_{c}\frac{{\bf r dr}}{r^3}.
=-\frac{\bf p}{c^2}\cdot I\oint_{c}\frac{{\bf r dr}}{r^3},
\label{pems}
\end{equation}
where $\bf p$ is the electric dipole moment of the final charge distribution.

The integral in Eq.~(\ref{pems}) can be written in terms of anti-symmetric and symmetric combinations of $\bf rdr$ as
\begin{equation}
{\bf P}_{\rm EM}=-\frac{I\bf p}{c^2}\cdot\oint_{c}\left\{\frac{{\bf [r(dr)-(dr)r]}}{2r^3}
+\frac{{\bf [r(dr)+(dr)r]}}{2r^3}\right\}.
\label{pems2}
\end{equation}
The symmetric integral vanishes because the numerator is a perfect differential, and the denominator $r^3$ is constant for the circular current loops of the solenoid.
The remaining integral  gives
\begin{eqnarray}
{\bf P}_{\rm EM}&=&-\frac{I}{c^2}\oint_{c}\frac{{\bf [(p\cdot r)dr-(p\cdot dr)r]}}{2r^3}\nonumber\\
&=&-\frac{\bf p}{c^2}\times I\oint_{c}\frac{{\bf dr\times r}}{2r^3}\nonumber\\
&=&\frac{\bf B\times p}{2c}.
\label{pems3}
\end{eqnarray}

This result agrees, as it must, with our other calculations of the EM momentum of an electric dipole in a uniform magnetic field.
It holds for the dipole moment of any static charge distribution in a uniform magnetic field. In particular, it holds for a parallel plate capacitor in the uniform magnetic field at the center of a solenoid, provided that the dimensions of the capacitor are much smaller than the radius of the solenoid.  
Reference \cite{brbg} found a corresponding result for ${\bf P}_{\rm EM}$ of a capacitor
without the factor 1/2, because they did not include the force on the solenoid while the electric dipole moment was being formed.
They introduced hidden momentum to account for the difference between their result and other calculations of
${\bf P}_{\rm EM}$ for this case.  But we see that including the force on the solenoid gives a result that is consistent with other calculations, with no need or room for hidden momentum.

\section{Center of energy theorem}

In this section we analyze the `center of energy theorem' that states\cite{dgl} ``If the center of energy of a closed system is at rest, then its total momentum is zero."
Application of this theorem leads to the introduction of hidden momentum to achieve  zero total momentum by canceling the EM momentum.
Reference\cite{dgl} refers to proofs of the center of energy theorem  by Coleman and Van Vleck\cite{cvv}, and Calkin\cite{calkin}, and we consider those proofs below.

Coleman and Van Vleck, and Calkin base their proof of the theorem on the vanishing divergence of the stress energy-momentum tensor:
\begin{equation}
\frac{\partial T^{\mu\nu}}{\partial x_\mu}=0.
\end{equation}
For the $T^{\mu 0}$ component
this reduces to
\begin{equation}
{\partial_t u}+{\grad\cdot\cal S}=0,
\label{eq:sud}
\end{equation}
where $\cal S$ is the energy flux vector, and $u$ is the energy density.
They say that in the static case $\partial_t u=0$, and conclude that $\grad\cdot{\cal S}=0$.  They then use this to prove the center of energy theorem.
Following Calkin, the proof first equates the momentum density $\bf g$ to ${\cal S}/c^2$, and then
\begin{eqnarray}
{\bf P}&=&\frac{1}{ c^2}\int{\cal S}d^3r\nonumber\\
&=&\frac{1}{ c^2}\int({\cal S}\cdot\grad){\bf r} d^3r\nonumber\\
&=&\frac{1}{c^2}\int[\grad\cdot({\bf r}{\cal S}) -{\bf r}(\grad\cdot{\cal S})]d^3r\nonumber\\
&=&\frac{1}{c^2}\oint{\bf dS\cdot r}{\cal S}+\int{\bf r}(\partial_t u) d^3r.
\label{eq:calk}
\end{eqnarray}
The surface integral vanishes if $\cal S$ vanishes fast enough as
the surface is taken to infinity, while the volume integral vanishes if
$\partial_t u=0$.  This would prove the theorem that $\bf P=0$.

There are several things wrong with this approach.  While Eq.~(\ref{eq:sud}) may hold for mechanical systems, it is not complete for EM energy. In fact, 
Poynting's theorem for EM energy
\begin{equation}
-\int{\bf j\cdot E}d^3r=\int(\partial_t  u) d^3r+\int(\grad\cdot{\cal S})d^3r,
\label{eq:pt}
\end{equation}
has an extra term that contradicts Eq.~(\ref{eq:sud}).
This leads to an inhomogeneous energy continuity equation
\begin{equation}
{\partial_t u}+{\grad\cdot\cal S}=-{\bf j\cdot E},
\label{eq:sudi}
\end{equation}
in contrast to Eq.~(\ref{eq:sud}).
The inclusion of this term changes the result of Eq.~(\ref{eq:calk}) to
\begin{equation}
{\bf P}_{\rm EM}=\frac{1}{c^2}\int{\bf r j\cdot E}d^3r,
\label{eq:c2}
\end{equation}
instead of equaling zero.  This refutes the supposed proofs of the center of energy theorem in Refs.~\cite{cvv} and \cite{calkin}.

The negative integral of $\bf j\cdot E$ in Eq.~(\ref{eq:pt}) corresponds to the rate of energy input to the EM fields, but the integral of
$\bf r j\cdot E$ has a different significance in Eq.~(\ref{eq:c2}).
In fact, it is just the EM momentum, ${\bf P}_{\bf rjE}$ as given in Eq.~(\ref{eq:pemrje}) above.
We see that objects at rest can have total momentum, because Poynting's theorem shows that
EM fields do not satisfy Eq.~(\ref{eq:sud}), and the center of energy theorem doesn't hold.
 
\section{Hidden momentum}
  
Because of considerable faith in the center of energy theorem, there have been numerous proposals for hidden momentum.  The hidden momentum has
no other purpose or physical manifestation than 
to cancel EM momentum, resulting in the zero momentum required by the center of energy theorem.  
There is no way to detect hidden momentum, which explains its name.  

Reference \cite{dgl} lists 43 references for hidden momentum, including two textbooks\cite{dj,dg}.
Several mechanisms are proposed in Ref.~\cite{dgl} for achieving hidden momentum in a current loop in the presence of an external constant electric field.
The first proposal utilizes a model of a current loop composed of conduction charges in a wire\cite{brbg}.  The charges accelerate on one side of the loop and decelerate
on the opposite side in response to the external electric field.
This would cause an imbalance in the relativistic mechanical momentum of the charges in the wire that would lead to hidden momentum.  This model relies on the external electric field penetrating the wire to do its hidden job.  

However, that is not what happens to a current in an external electric field.  With no current, the free charges rearrange themselves, forming a surface charge that prevents the electric field from entering the conductor. 
When a current is put into the wire, the displaced surface charge distribution still prevents the external electric field from having any effect on the current.  
Thus this hidden momentum mechanism, as proposed in Ref.~\cite{brbg}, is absent even for the artificial model of electric current they propose.   

Several  references\cite{dgl,wf,vh,calkin} recognize correctly that the fact that a static external electric field cannot influence the current in a  conductor means that such a current cannot have hidden momentum.  But then  they claim that this would also lead to no EM momentum, thus preserving the center of energy theorem. 
 However, we have shown in our derivation of Eq.~(\ref{eq:edip}) that any magnetic moment, even that of a conducting wire, would give the ${\bf P}_{\rm EM}$ of Eq.~(\ref{eq:edip}).
The induced charges (We could designate them as $\rho_2$.) that cancel the effect of $\rho_1$ inside the wire
do not enter the calculation of  
${\bf P}_{\rm EM}$ because they would give self-acting forces that cannot affect the momentum of the wire.  
Only the electric field (${\bf E}_1$) of the external charge distribution ($\rho_1$) affects the momentum of the current in the wire.
That is why we used distinct charge-current distributions, $\rho_1$ and ${\bf j}_2$, in our starting equation (\ref{eq:pem1}). 
Thus a current in a wire does lead to EM momentum, in contradiction of the center of energy theorem.

There will be an ohmic electric field in the wire, given by 
\begin{equation}
\bf E=j/\sigma,
\label{ohm}
\end{equation}
which does not affect the EM momentum.
In fact, Eq.~(\ref{ohm}) shows that if the current density is constant (as required for this mechanism of hidden momentum) then any electric field in the wire must be constant along its length. This then forbids any electric field that could introduce differential acceleration around the current loop. Thus the above mechanism for hidden momentum contradicts itself.

It has also been proposed\cite{vh} that a static electric field could accelerate charges moving in a non-conducting tube, and thus lead to hidden momentum.
However, this will not work either because any external static electric field that could accelerate the charges in the current would first rearrange them to keep any static electric field out of the current.  Thus, any current, no matter how contrived, would have no hidden momentum in an electric field, but the combined current-charge distribution would have EM momentum, contradicting the center of energy theorem.

The original proposal\cite{sj} for hidden momentum was for moving charges in two oppositely charged rotating disks, with the charges fixed in each disk.
A similar mechanism was later proposed for charges carried by an incompressible fluid enclosed in a nonconducting tube\cite{vh}.
In each case, the charges cannot be moved around to exclude an external electric field.
This eliminates the exclusion of the external electric field, but then there can be no varying acceleration as in the first proposal.  

In the presence of an external electric field, this proposal does appear to lead to a momentum that is equal and opposite to the EM momentum. 
Ref.~\cite{vh} achieves this by writing the continuity equation for energy in matter as\cite{hi}
\begin{equation}
{\partial_t u}+{\grad\cdot\cal S}=+{\bf j\cdot E}.
\label{eq:sudm}
\end{equation}
The inhomogeneous term in this equation is the negative of that in the corresponding equation (\ref{eq:sudi}) for energy in the EM field.
This is because any energy that leaves the matter goes into the EM field.
Using this equation leads to Eq.~(6) of ref.~\cite{vh}
\begin{equation}
{\bf P}_{\rm Hidden}=
-\frac{1}{c^2}\int{\bf rj}\cdot{\bf E}d^3r
\label{eq:pmat}
\end{equation}
for the presumed hidden momentum of the incompressible fluid of Ref.~\cite{vh} or the rotating disks of Ref.~\cite{cvv}.
This momentum is equal and opposite to the EM momentum  as given by our Eq.~({\ref{eq:pemrje}), but
it should not be added to the EM momentum so as to cancel it.  To do so would be to add the action and reaction forces of Newton's third law.  

The momentum in Eq.~(\ref{eq:pmat}) is not hidden momentum. 
It is the momentum that
would result from the Lorentz force on the charge distribution in our Eq.~(\ref{eq:fmat})
if there were no external force.  It is in fact the time integral of 
${\bf F}_{\rm Matter}$ in our Eqs.~(\ref{eq:fm}) and(\ref{eq:n3}).
It would be the momentum given to the charge distribution (not the current distribution) 
were it not for the action of the external force in Eq.~(5) that keeps the charge distribution at rest.
Although Refs.~\cite{vh} and \cite{cvv} consider the momentum in Eq.~(\ref{eq:pmat}) to be hidden momentum in the current distribution, it would actually be the mechanical momentum of the charge distribution if there were no external force to keep it at rest.
 
In order to keep the charge distribution at rest, the momentum in Eq.~(\ref{eq:pmat}) would be canceled by the time integral of the external force
(as we discussed in Sec~2), and is absent from the final state of the matter.  That is, Eq.~(6) of Ref. \cite{vh} should be  replaced by
\begin{equation}
{\bf P}_{\rm Matter}=
-\frac{1}{c^2}\int{\bf r(j}\cdot{\bf E})d^3r-\int{\bf F}_{\rm External}dt,
\end{equation}
where ${\bf F}_{\rm External}$ is the force that is required to keep the matter at rest, resulting in ${\bf P}_{\rm Matter}={\bf 0}$.
The momentum in Eq.~(\ref{eq:pmat}) is not hidden momentum, but
whatever it is called, it is zero because of the action of the external force.
There is no hidden momentum in this example.
 
We have seen that none of the proposed mechanisms lead to hidden momentum.
The presumed appearance of hidden momentum to cancel EM momentum recalls a parable from our youth.  
Look carefully at Eqs.~(2), (3), and (5) of this paper.
$d{\bf P}_{\rm EM}/dt$ is like the force on the cart (representing the EM field)
by the horse.  $d{\bf P}_{\rm Matter}/dt$ is like the force on the horse by the cart.
They are equal and opposite by Newton's third law, but act on different objects.
${\bf F}_{\rm External}$ is like the force of the ground on the horse's hooves.
If the horse's mass is negligible compared to the cart, then ${\bf F}_{\rm External}$ is equal and opposite to $d{\bf P}_{\rm Matter}/dt$ by Newton's second law\cite{rest}.
Although each pair of forces are equal and opposite, a free body diagram of the cart and the horse together shows that ${\bf F}_{\rm External}$ is the only external force.
This gives the combined system a total momentum equal to the time integral of 
${\bf F}_{\rm External}$, which is not zero.  This contradicts the `cart can't move' theorem.

\section{Conclusion}

Our conclusion is that the `center of energy' theorem does not apply to the EM momentum of a static charge-current distribution, and that `hidden momentum' is neither needed nor present in the charge-current distribution.
The external force needed to keep matter at rest during the creation of the charge-current distribution goes directly into EM momentum without moving any matter or hiding any momentum.


\begin{thebibliography}{9}
\bibitem{dgl} ``Resource Letter EM-1: Electromagnetic Momentum", D. J. Griffiths, Am.~J. Phys.~{\bf 80}, 7-18 (2012).
\bibitem{b1}An exception is ``Interaction of a Point Charge and a Magnet: Comments on `Hidden
Mechanical Momentum Due to Hidden Nonelectromagnetic Forces' ", T. H. Boyer, arXiv:0708.3367 (2007).
\bibitem{tdj}We don't use the alternative of using a current to produce the static charge distribution here, because that would result in two separate forces, one on each current distribution.  
\bibitem{brbg} ``Hidden momentum, field momentum, and electromagnetic impulse", D. Babson, S. P. Reynolds, R. Bjorkquist, and D, J. Griffiths,
Am. J. Phys.~{\bf 77}, 826-833 (2009).
\bibitem{wf} ``Examples of momentum distributions in the electromagnetic
field and in matter", W. H. Furry, Am. J. Phys.~{\bf  37}, 621-636 (1969).
\bibitem{vh} ``Hidden momentum of a relativistic fluid carrying current in an external electric
field",  V. Hnizdo, Am. J. Phys.~{\bf  65}, 92-94 (1997).
\bibitem{jf} {\it Classical Electromagnetism}, J. Franklin
(San Francisco: Addison Wesley, 2005)
\bibitem{dj}{\it Classical Electrodynamics} 3rd Edn, J. D. Jackson
(New York: John Wiley \& Sons, 1999)
\bibitem{cvv}``Origin of ‘hidden momentum forces’ on magnets", S. Coleman and J.
H. Van Vleck, Phys. Rev.~{\bf 171}, 1370-1375 (1968).
\bibitem{calkin}``Linear Momentum of the Source of a Static Electromagnetic Field", M. G. Calkin, Am.~J. Phys.~{\bf39}, 513-516 (1971).
\bibitem{dg}{\it Introduction to Electrodynamics}, 3rd Edn, D. J. Griffiths  (San Francisco: Addison Wesley, 1999)
\bibitem{sj}``Try simplest cases’ discovery of ‘hidden momentum’ forces on ‘magnetic currents’", W. Shockley and R. P. James,
 Phys. Rev. Lett.~{\bf 18}, 876-879 (1967).
\bibitem{hi}Although this equation contradicts Eq.\ (\ref{eq:sud}), which is used to prove the center of energy theorem, 
it is used here to preserve the center of energy theorem.
\bibitem{rest}For a real horse, ${\bf F}_{\rm External}$ would be greater than ${\bf F}_{\rm Matter}$.  For the example in Sec.~1, the charge distribution is kept at rest, so 
${\bf F}_{\rm External}$ and ${\bf F}_{\rm Matter}$ are equal and opposite by Newton's first law.
\end{thebibliography}
\end{document}